\documentclass[useAMS,usenatbib]{mn2e}
\usepackage{natbib}
\usepackage{url,threeparttable}
\usepackage{amssymb,amsmath}
\usepackage{txfonts,graphicx}

\newcommand\aj{{AJ}}%
\newcommand\araa{{ARA\&A}}%
\newcommand\apj{{ApJ}}%
\newcommand\apjl{{ApJ}}%
\newcommand\apjs{{ApJS}}%
\newcommand\aap{{A\&A}}%
\newcommand\aaps{{A\&AS}}%
\newcommand\mnras{{MNRAS}}%
\newcommand\pasp{{PASP}}%
\catcode`\@=11
\def\gapprox{\mathrel{\mathpalette\@versim>}}
\def\lapprox{\mathrel{\mathpalette\@versim<}}
\def\@versim#1#2{\lower2.45pt\vbox{\baselineskip0pt\lineskip0.9pt
      \ialign{$\m@th#1\hfil##\hfil$\crcr#2\crcr\sim\crcr}}}
\catcode`\@=12
 
\newcommand{\hi}{H\,{\sc i}}

\newcommand{\ha}{\ensuremath{\mbox{H}\alpha}}
\newcommand{\nii}{N\,{\sc ii}}

\newcommand{\mum}{\ensuremath{\,\mu\mbox{m}}}

\newcommand{\asec}{\ensuremath{^{\prime\prime}}}
\newcommand{\amin}{\ensuremath{^{\prime}}}
\title[The warm dust and old stars of M31's Bulge]{The heating of dust by old stellar populations in the Bulge of M31}
\author[B. Groves et al.]{Brent Groves\thanks{brent@mpia.de}$^{1}$,
  Oliver Krause $^{1}$,  Karin Sandstrom$^{1}$, Anika
  Schmiedeke$^{1, 2}$, 
 \newauthor
 Adam Leroy$^3$, Hendrik Linz$^{1}$, Maria Kapala$^1$, Hans-Walter Rix$^{1}$, 
\newauthor
Eva Schinnerer$^{1}$,  Fatemeh Tabatabaei$^{1}$, Fabian
Walter$^{1}$ and Elisabete da Cunha$^{1}$
\vspace*{6pt}\\
$^{1}$Max Planck Institute for Astronomy, K\"{o}nigstuhl 17, D-69117
Heidelberg, Germany\\
$^{2}$Universit\"at zu K\"oln, Z\"ulpicher Strasse 77, 50937 K\"oln, Germany\\
$^{3}$National Radio Astronomy Observatory, Charlottesville, VA 22903, USA}
\begin{document}
\maketitle
\begin{abstract}
We use new Herschel multi-band imaging of the Andromeda galaxy to
analyze how dust heating occurs in the central regions of galaxy
spheroids that are essentially devoid of young stars.  We construct a
dust temperature map of M31 through fitting modified blackbody SEDs to
the Herschel data, and find that the temperature within 2 kpc rises
strongly from the mean value in the disk of $17\pm1$\,K to $\sim35$\,K
at the centre.  UV to near-IR imaging of the central few kpc shows
directly the absence of young stellar populations, delineates the
radial profile of the stellar density, and demonstrates that even the
near-UV dust extinction is optically thin in M31's bulge.  This allows
the direct calculation of the stellar radiation heating in the bulge,
$U_{\ast}(r)$, as a function of radius.  The increasing temperature
profile in the centre matches that expected from the stellar heating,
i.e. that the dust heating and cooling rates track each other over
nearly two orders of magnitude in $U_{\ast}$. The modelled dust heating is
in excess of the observed dust temperatures, suggesting that it is
more than sufficient to explain the observed IR emission. Together with the
wavelength dependent absorption cross section of the dust, this
demonstrates directly that it is the optical, not UV, radiation that
sets the heating rate. This analysis shows that neither young stellar
populations nor stellar near-UV radiation are necessary to heat dust
to warm temperatures in galaxy spheroids. Rather, it is the high
densities of Gyr-old stellar populations that provide a sufficiently
strong diffuse radiation field to heat the dust. To the extent which
these results pertain to the tenuous dust found in the centres of
early-type galaxies remains yet to be explored.
\end{abstract}

\section{Introduction}\label{sec:intro}

As the nearest, massive galaxy, Andromeda
(M31, NGC\,224)  has offered a unique insight into the properties of
galaxies. It provides the perfect stepping stone between the well-resolved
interstellar medium (ISM) and stellar populations within our own Galaxy, and the integrated
properties of more distant galaxies. Due to its proximity ($\sim
780$kpc), Andromeda offers more than
a resolved example of an early-type spiral galaxy, but can also be used
to explore the interaction of the ISM and stars in early-type galaxies (ETGs) through
its large, visually dominant bulge.

Andromeda's bulge dominates the stellar luminosity and mass within the
central 1.5 kpc \citep[$R_{\rm eff}({\rm bulge})\sim0.5-1.1$\,kpc;][]{Courteau11}, with the disk
dominating beyond this.  
The bulge contributes $\sim$30\% of the total stellar mass
and luminosity in M31, and it is clearly the highest surface
brightness feature at UV -- NIR wavelengths \citep{Geehan06,Courteau11}.
While the integrated colours and luminosity of M31 may place the
galaxy in the ``green valley'' \citep{Mutch11}, the optical colours of
the bulge are red, with a $B-V \approx 0.9$ 
to 1.0 \citep{Walterbos87a}, placing it securely in the range of the
red-sequence, where most early-type galaxies are found. \citet{Oke68}
even used the spectral energy distribution of the centre of M31 as a
representative for the average giant elliptical galaxy when
determining $K$-corrections.

This ``early-type'' nature of the bulge is supported by the old
mean stellar age determined for the central region of M31. Resolved
star colour-magnitude diagrams created by ground-based,
adaptive-optics, NIR imaging reveal a population dominated
by stars greater than 6 Gyr \citep{Davidge05,Olsen06}, while
single-stellar population fits to absorption-line indices from
slit-spectroscopy of the bulge \citep{Saglia10} find that the bulge of
M31 is uniformly old ($\ge12$ Gyr, excluding the central
arcsecs). High-resolution, individual star photometry from the
Pan-Chromatic Hubble Andromeda Treasury  survey
\citep[PHAT;][]{Dalcanton12} find no population of
young-stellar sources, with the UV-light dominated by evolved stars
such as post-AGB stars \citep{Rosenfield12}.

Given this lack of young stars and star-formation, it is unsurprising
that the bulge of M31 is also extremely gas 
poor. Little to no CO is detected in the CO(1-0) map of
\citet{Nieten06} down to low surface brightnesses. CO is
detected in the central part of M31 when deeper observations of small high
attenuation regions are made \citep{Melchior00,Melchior11}, however
the attenuation in these regions is still relatively low ($A_{\rm B}
<0.3$) and the covering fraction of these regions is very small \citep{Melchior00},
meaning diffuse, low attenuation dust (and presumably diffuse gas) is
more characteristic of the bulge. Compounding this is a low 
CO-H$_2$ ratio \citep{Leroy11}, and a low total HI column
\citep{Braun09} giving a total cool gas mass in the centre of only a few $10^6
M_{\odot}$ ($\sim0.02$\% of the bulge stellar mass), likely dominated
by molecular gas \citep[e.g.][]{Li09,Melchior11}. 

This low gas fraction makes the M31 bulge more gas-poor than many
ETGs, as recent work with the ATLAS3D sample \citep{Cappellari11} has
demonstrated. \citet{Young11} observed CO(1-0) in $\sim$22\% of the
sample of nearby early-type galaxies, giving corresponding gas masses
of $M(\rm {H}_2) > 10^{7}M_{\odot}$ (for a sample with a median
stellar mass of $M_{\star}=3\times10^{10}M_{\odot}$). Similarly,
studies of nearby field ellipticals and lenticulars have found atomic
gas in a large fraction of these ($\sim70$\%), with this \hi\ emission
generally associated with ionized gas emission
\citep{Morganti06}. Interestingly, the molecular gas in early-type
galaxies is not always associated with star formation. Optical
colours and emission-line ratios indicate other forms of heating in
the ISM of some early-types, and only low levels, if any, of star formation
\citep{Crocker11}.

Given the very low level of star formation, low attenuation, and small
amount of gas at the centre of M31, little dust emission was expected
in the central kiloparsecs of M31. Yet when this region was examined
in the far-IR with IRAS \citep{Habing84}, ISO \citep{Haas98}, and
\emph{Spitzer} \citep{Gordon06}, emission at wavelengths greater than
60\,\mum\ was clearly seen,
indicating the presence of warm dust. In nearby ETGs, low-levels of
warm dust are also being detected, with \citet{Smith11} finding dust
emission in 24\% of ellipticals and 62\% of S0s in the \emph{Herschel}
Reference survey. While $M_{\rm dust}/M_*$ is far lower in spheroids
than in disks, the dust appears to be warmer on average than that found in
later-type galaxies.  A similar result was found using \emph{Herschel}
in nearby early-type spirals similar to M31 by \citet{Engelbracht10},
with the mean dust temperature of the bulges consistently hotter than
the disks in these galaxies. Relatively higher dust temperatures were
also found by \citet{Rowlands11} in the \emph{Herschel}-ATLAS survey
of more distant ellipticals. Hence dust, when detected, tends to be in
a warmer state in spheroids than that found across the disks of
later-type galaxies.

Given the high density of stars in bulges, stellar heating is the likeliest
explanation for the observed warmer dust temperatures in ETGs. Yet the
old stellar ages found in ETGs, and especially in the centre of M31, argue against the standard view
that the IR luminosity and warm dust is a direct tracer of star formation
\citep[e.g.][]{Kennicutt98}. Based on the IRAS observations and the
extremely weak UV observed using the Astronomical Netherlands
Satellite \citep[ANS;][]{Coleman80},
\citet{Habing84} put forward the argument that it is the high density of
late-type giant stars that provide the strong enough radiation field to heat
the dust. 
Based on \emph{Herschel} PACS and SPIRE maps of
M31 with unprecedented resolution (Krause et al., in prep.), we follow
\citet{Habing84} and use this high resolution to
demonstrate how and by what stellar 
populations dust is heated in the bulge of this galaxy, and, by proxy,
in the spheroids of other inactive early-type galaxies. 

We briefly review the \emph{Herschel} data and reduction in section \ref{sec:data},
and discuss the FIR geometry and integrated properties of the M31 bulge in
section \ref{sec:FIR}.  In section \ref{sec:bulgeheat} we determine
and discuss the
 mean dust temperature across M31 and in the centre, and determine the
 heating mechanism for the rise in dust temperature in the centre.
We finish with the discussion and summary in sections
\ref{sec:disc} and \ref{sec:summ}. The global properties of M31 we assume throughout this
paper are listed in Table \ref{tab:m31prop}.

\begin{table}
 \caption{M31 positional data$^{\rm a}$}\label{tab:m31prop}
 \smallskip
  \begin{threeparttable}[\hsize]
      \begin{tabular}{ll}
      \hline
      M31 nucleus position\tnote{b}  & $00^{h}42^{m}44.35^{s}$\\
      ~(J2000)&$+41^{\circ}16\amin08.60\asec$\\
      Position angle of major axis & $37.7^{\circ}$\\ 
      Inclination & $75^{\circ}$\\
      Distance\tnote{c} & $780\pm40$ kpc\tnote{d}\\
      Distance modulus & 24.46\\
      \hline
     \end{tabular}
     \begin{tablenotes}
       \item[a] Based on NED data and references where given.
       \item[b] \citep{Evans10}, and verified in \emph{Spitzer} IRAC 3.6\mum\ image.
       \item[c] \citet{Stanek98,Rich05}, see also NED for other determinations.
       \item[d] $1\amin=227\pm12$pc along major axis
     \end{tablenotes}
  \end{threeparttable}
\end{table}

\section{Data and reduction}\label{sec:data}

The M31 \emph{Herschel} data and their reduction are described and
discussed in detail in Krause et al. ~(in prep.), so we only briefly
review the data here. 

M31 was imaged in all 6 \emph{Herschel} photometric  bands (PACS 70, 100, and
160\mum, and SPIRE 250, 350, and 500\mum) in slow parallel mode
for a total time of $\sim24$ hours. The images extend for $\sim3^{\circ}\times1^{\circ}$,
centred slightly off-nucleus (at the P.A. given in Table
\ref{tab:m31prop}), covering the main stellar disk of Andromeda,
including the 10kpc ring.

All images were reduced to level one using HIPE version 6.0, and then
used SCANAMORPHOS v12.0 \citep{Roussel12} to produce the final images.
As HIPE v6.0 was used, we converted the PACS images from flight model (FM), 5 to FM, 6 by dividing
by the factors listed in PACS Photometer Point Source Flux
Calibration Report v1.0, which are 1.119 (70\mum), 1.151 (100\mum),
and 1.174 (160\mum) for the three PACS photometer bands, respectively. 
All images were converted to MJy\,sr$^{-1}$. Based on SPIRE Observers'
manual v2.4\footnote{ \url{http://herschel.esac.esa.int/Docs/SPIRE/html/spire_om.html}},
we used beam areas of 423$\square\asec$ (250\mum), 
751$\square\asec$ (350\mum), and 1587$\square\asec$ (500\mum) for
the three SPIRE bands, respectively, to convert from Jy\,beam$^{-1}$ to MJy\,sr$^{-1}$.

The mean FWHM of the PACS and SPIRE Point Response Functions/beams are; $\sim5.6\asec$
(70\mum), $\sim6.8\asec$ (100\mum), $\sim11.4\asec$  (160\mum), 18.2\asec\
(250\mum), 24.9\asec\ (350\mum), and 36.3\asec\ (500\mum),
respectively, for the 20\asec/s scans used here (for full details of
the correct point spread functions and beams, see the PACS\footnote{
  \url{http://herschel.esac.esa.int/Docs/PACS/html/pacs_om.html}} and
SPIRE$^1$ Observer manuals).  The actual images use pixels sizes of 1\asec\ for
all PACS bands and 6\asec, 10\asec, and 14\asec, for the respective
SPIRE bands. When comparing bands, we convert all bands to the lowest
resolution and pixel size using the convolution kernels provided by
\citet{Aniano11}.

For all bands we have measured the noise and background as discussed
in Krause et al. (in prep.). The `background' in the images is a
combination of both Galactic foreground and background galaxies, and
is significant only in the SPIRE images. For simplicity, a uniform
background in each band was assumed and subtracted from the image.

The measured
noise for each band in the original images is; 3.17 MJy\,sr$^{-1}$ (70\mum),
3.23 MJy\,sr$^{-1}$ (100\mum), 2.29 MJy\,sr$^{-1}$ (160\mum), 0.68 MJy\,sr$^{-1}$ (250\mum),
0.46 MJy\,sr$^{-1}$ (350\mum), and 0.20 MJy\,sr$^{-1}$ (500\mum). When convolved and
binned to lower resolutions, the noise per resolution element in each
band is obviously reduced, with the effective resolution driven by the
longest wavelength (e.g.~36.3\asec\ FWHM and 14\asec\ pixels at the
500\mum\ resolution).

\section{The bulge of M31: Stars, dust, and gas}\label{sec:FIR}
\subsection{The Far-IR emission in Andromeda}
\begin{figure*}
\includegraphics[width=\hsize]{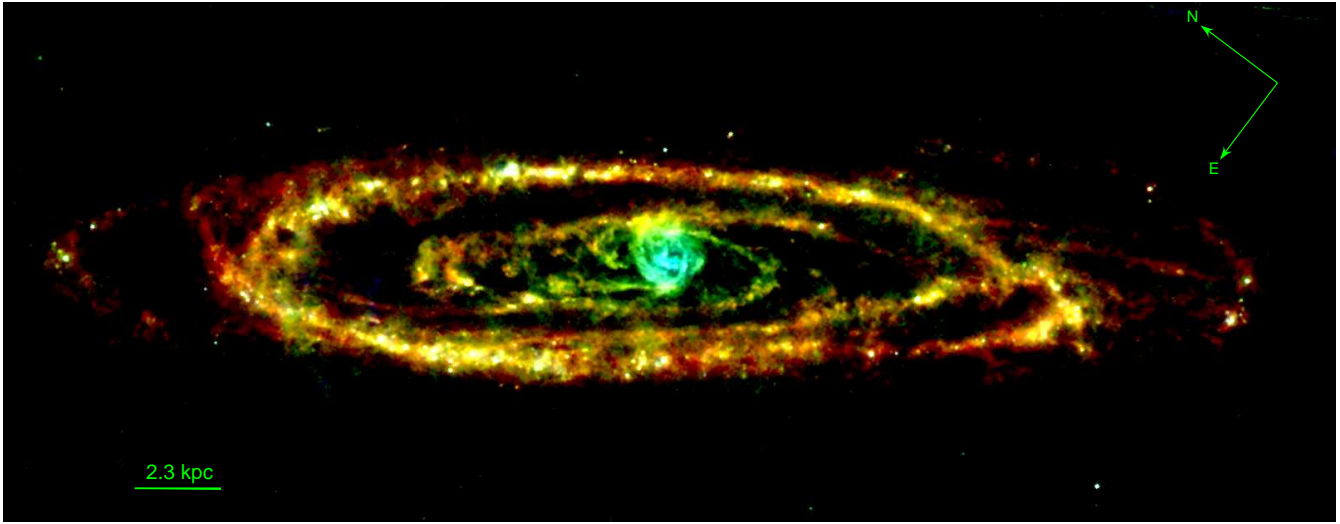}
\caption{Far-infrared image of Andromeda using the Herschel bands
  showing PACS 70\mum\ (\emph{blue}), PACS 100\mum\ (\emph{green}), 
  and SPIRE 250 \mum\ (\emph{red}), all convolved to
  SPIRE 250\mum\ resolution (from Krause et al., in prep.). All three bands have the same square root
  scaling from 10 MJy\,sr$^{-1}$ to 150  MJy\,sr$^{-1}$.
  The angular scale of the image is shown by the 2.3\,kpc (10\amin) bar in the
  lower left. }\label{fig:m31image} 
\end{figure*}

In Figure \ref{fig:m31image} we show a RGB image of the full disk of
Andromeda using the PACS 70\mum, PACS 100\mum, 
and SPIRE 250 \mum\ bands.
One of the things that stands out in this image, apart from the dusty
ring, is the blue centre of the IR map, indicating a distinct rise in dust temperature
towards the centre of M31.

The central kiloparsec bulge region (hereafter referring to a circular aperture of radius
4.405\amin (1\,kpc), centred on the stellar peak defined in Table
\ref{tab:m31prop}) is relatively weak in the SPIRE 250\mum\ band
compared to the extended disk and star forming ring, but shows
a relative excess in the PACS 70\mum\ band.
The far-IR emission in this central region appears more circular and
shows a spiral like pattern within.

\subsection{The correlation of dust and gas}\label{sec:dang}
\begin{figure*}
\includegraphics[width=0.48\hsize]{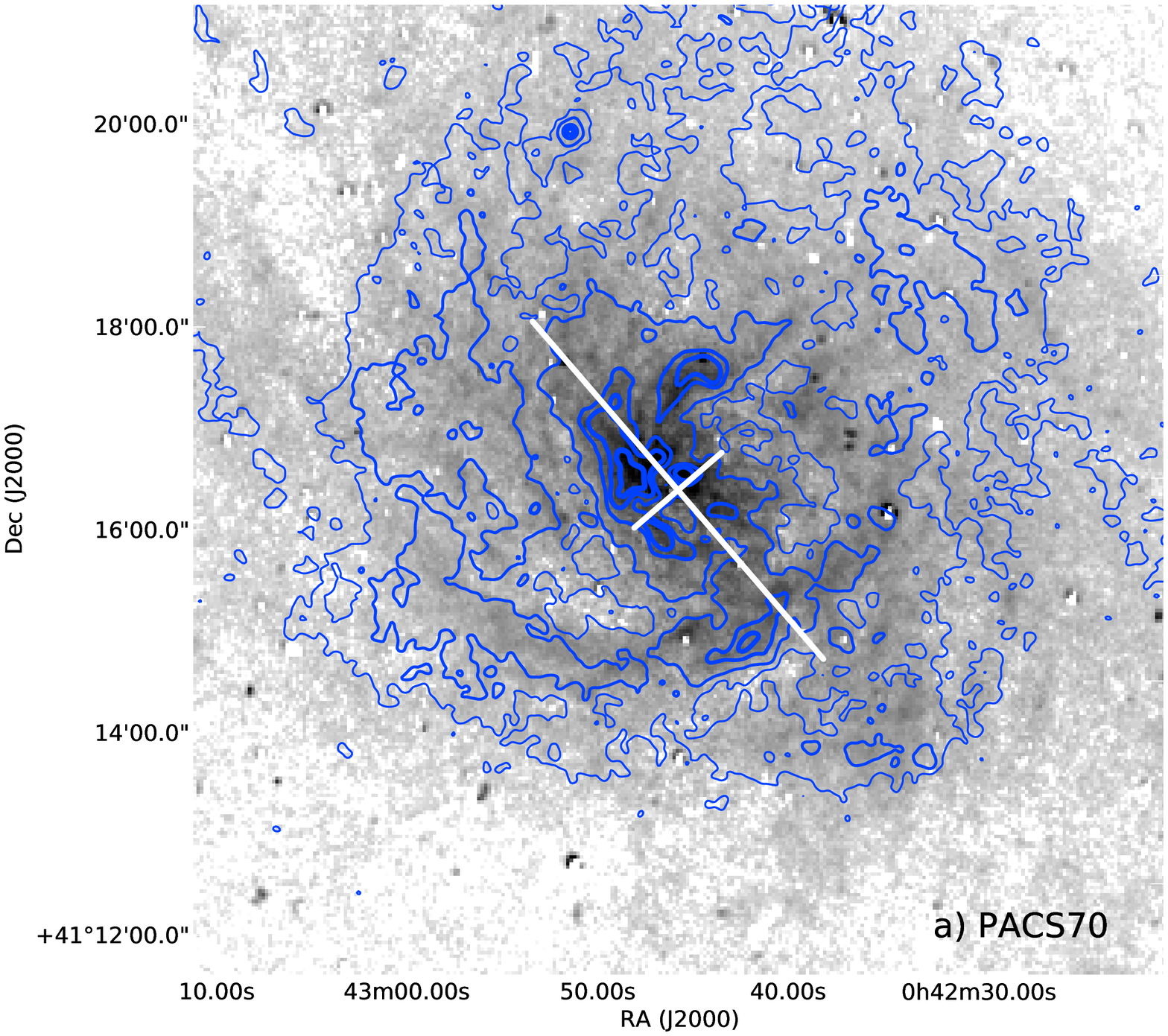}
\includegraphics[width=0.48\hsize]{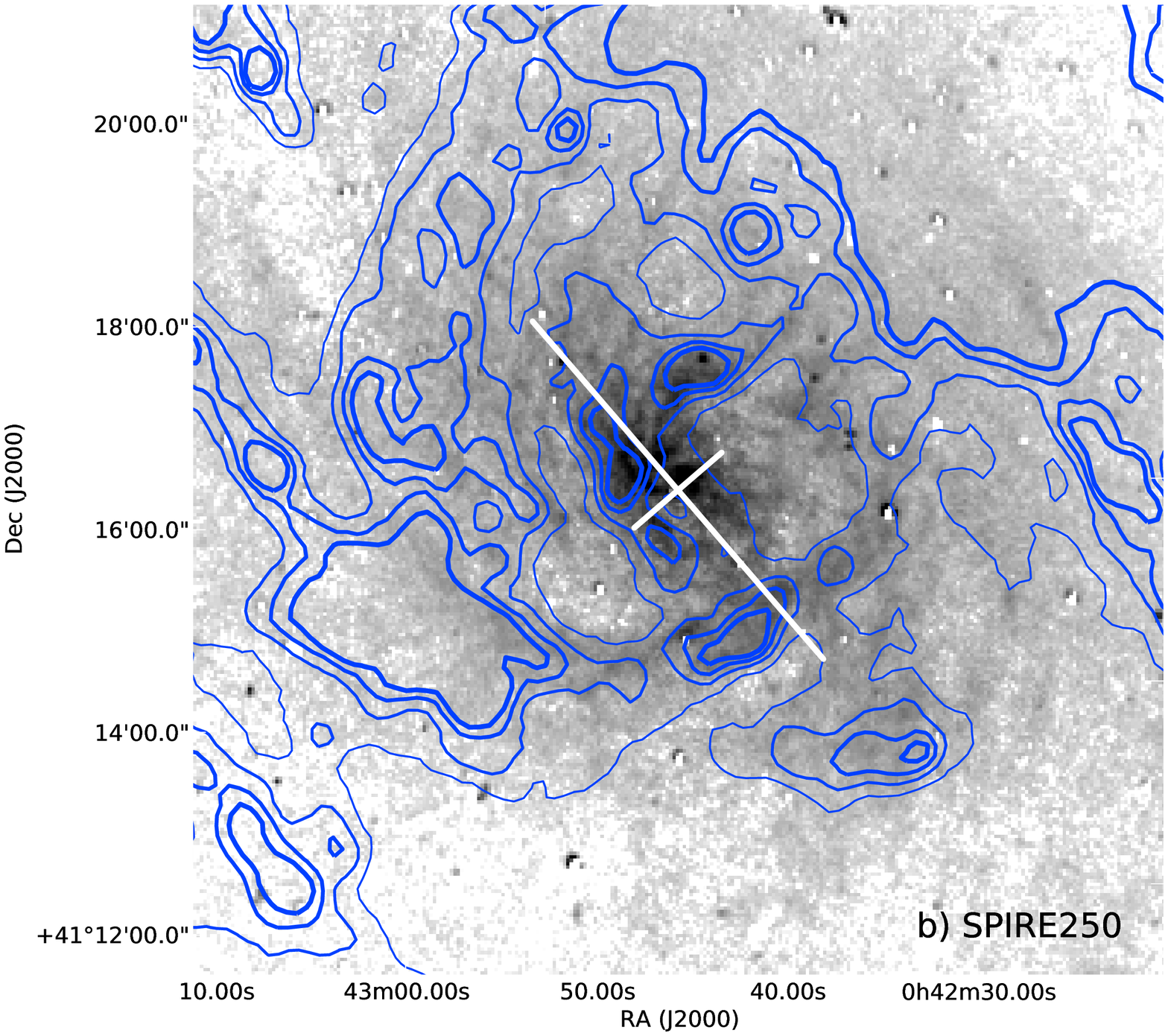}
\caption{Continuum-subtracted H$\alpha+$[N{\sc ii}] narrow band image
  of the central kiloparsecs of M31 (\emph{greyscale}) from
  \citet{Devereux94} overlaid with contours from PACS 70\mum\ (\emph{left})
  and SPIRE 250\mum\ (\emph{right}) imaging. In both images the \ha\
  has a square root stretch, and the centre is marked by a cross of 1
  kpc length along both the major (4.4\amin) and minor axis (1.14\amin), assuming the
  inclination of M31 given in Table \ref{tab:m31prop}. For the PACS
  70\mum\ contours, the levels are 15, 30, 70, and 100 MJy\,sr$^{-1}$, while for
  the SPIRE 250\mum\ contours, the levels are 10, 15, 20, and 25
  MJy\,sr$^{-1}$, each indicated by an increasing thickness.
}\label{fig:m31ha_IR}
\end{figure*}

The spiral pattern can be seen in finer detail with Figure \ref{fig:m31ha_IR},  where we show
the PACS 70\mum\ and SPIRE 250\mum\ contours overlaid on the \ha\
image of the central kiloparsecs from \citet{Devereux94}.
As demonstrated by \citet{Li09} using \emph{Spitzer} data
(particularly their Figures 6 and 8), the IR emission follows well the
morphology of the \ha\ emission. The correspondence is not
linear between the \ha\ and the two \emph{Herschel} bands, with offsets 
between the \ha\ peaks and IR peaks, but in general the same barred
spiral pattern is seen. The two \emph{Herschel} bands match very well
with each other within the inner 0.5 kpc (2.2\amin) radius, but outside this the correspondence
becomes weaker as the dust becomes cooler, and the PACS 70\mum\
emission merges with the noise. To the NW and SE of the image in
Figure \ref{fig:m31ha_IR}b) we can see the contribution of the disk
appearing at the SPIRE 250\mum\ wavelengths.

The correspondence of the dust and ionized gas emission together
suggest that the dust is associated with the gas and also likely to be
in the same lower inclination thin disk spiral
\citep{Jacoby85,Ciardullo88}. The offset between the peaks in the \ha\ 
and the IR emission indicates the presence of cooler, denser gas associated with the dust, that
is still too low in column density to be observed in \hi\ or CO, with
the weak and non-detections in CO of several higher
attenuation regions in the centre by \citet{Melchior00} and \citet{Melchior11}
supporting this. As the gas emission indicates that most of the gas is
in a diffuse state, and attenuation maps show an, on average,
extremely low attenuation, it is
clear that the dust is likely to be optically thin to a significant amount of the
radiation from the bulge stars.

\subsection{The spectral energy distribution of the bulge}\label{sec:SED}
\begin{figure}
\includegraphics[width=\hsize]{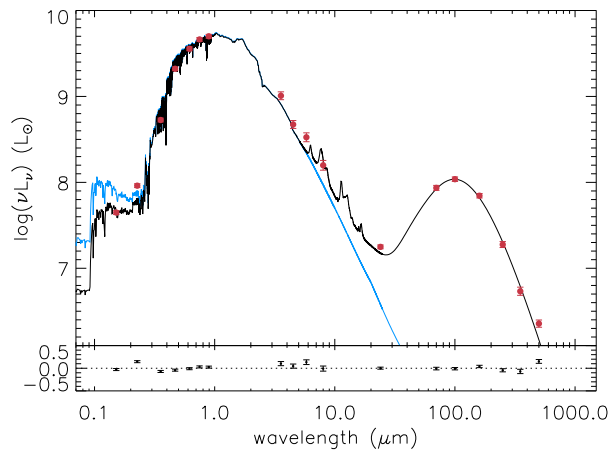}
\caption{The spectral energy distribution (SED) of the central kpc region of
  Andromeda, measured within a circle of 1kpc (4.405\amin) radius
from the centre. The fluxes (GALEX FUV and
 NUV, SDSS $u$, $g$, $r$, $i$, and $z$, \emph{Spitzer} IRAC and
 MIPS24, and all six \emph{Herschel} bands) within this aperture are marked by the red
 points, with the error bars including calibration, sky and noise
 uncertainties. The black curve shows the best fit model
SED from the SED fitting code
MAGPHYS \citep{daCunha08}, while the blue curve shows the
corresponding unattenuated stellar SED from the same model. Below the model SED,  the
residuals between the observed data and models are shown, with
$|\log(L_{\rm obs}/L_{\rm mod})|<0.1$ dex for all wavelengths except
for SPIRE500\mum, where the observed flux is under predicted by
$\sim$30\%.
}\label{fig:m31sed}
\end{figure}

The low optical depth in the centre of M31 is also made clear by the
full UV--IR SED in the central kiloparsec, which is illustrated in
Figure \ref{fig:m31sed}, drawing on data from GALEX
\citep{GildePaz06}, SDSS \citep{Aihara11}, \emph{Spitzer} IRAC
\citep{Barmby06} and MIPS \citep{Gordon06}, and now
\emph{Herschel}. Overlaid on the observed data is the best fit from
the SED fitting code MAGPHYS\footnote{\url{www.iap.fr/magphys}}
\citep{daCunha08} (black curve) and its 
implied unattenuated stellar emission (blue curve). For this model fit
we created a new stellar library using the original \citet{BC03} code
and the same star formation histories as in \citet{daCunha08} but
excluding all models which were formed less than $6$\,Gyrs ago
($t_{\rm g}< 6$\,Gyr, see \citet{daCunha08}, specifically \S 3.1.1), based on previous estimates for the
mean stellar age of the bulge of M31 \citep[see section \ref{sec:intro},
and e.g. ][]{Davidge05,Olsen06}.
The MAGPHYS model fit returns a stellar mass within this 1kpc radius of $\log
(M_{\star}/M_{\odot})=10.01\pm0.01$, and an unattenuated stellar
luminosity of  $\sim10^{9.9}\,L_{\odot}$. The dust luminosity of the M31 centre is
$\sim10^{8.3\,}L_{\odot}$, with MAGPHYS determining a total dust mass of
only $\log (M_{\rm dust}/M_{\odot}) = 5.17 \pm 0.05$.

As stated in the introduction, the $10^{10}M_{\odot}$ stellar mass of
the bulge makes it a significant fraction ($\sim 30$\%) of the total
stellar mass of M31 \citep[][]{Geehan06}.
The optical--UV colours clearly 
reveal the old stellar age of the bulge, with NUV-$r\approx5.0$. 
These colours lead to an
extremely low estimated specific star
formation rate from the model, with sSFR$<0.01 {\rm Gyr}^{-1}$ (SFR $<
10^{-2}M_{\odot}$), in agreement with 
the lack of young stars observed in the bulge of M31.
\citep{Davidge05, Olsen06, Rosenfield12}. 

It should be noted that the values here
for the dust mass and luminosity in the central kiloparsecs are likely
upper limits for the dust in the bulge, as even at these small radii,
some fraction of the IR emission still arises from the disk in the
foreground and background due to the high inclination of Andromeda. 
This contribution of the disk infrared emission may also explain the
weak UV ($<0.25\mum$) excess observed in the unattenuated spectrum of
the best fit MAGPHYS model relative to the observed SED.  
To explain the observed IR SED, the MAGPHYS model finds a small component of
buried young stars necessary. This component is small (as demonstrated
by the low sSFR), and cannot be associated with the bulge (as the high-resolution
observations have shown), but still contributes significantly to the
unattenuated UV.

The total attenuation in the diffuse ISM is extremely low, as can be
seen in the differences between the black and blue curves, with the
model finding an attenuation in the diffuse ISM of only $\tau_{\rm
  V}\sim 0.03$.
The low value of attenuation measured from the SED matches that
observed in $B$-band attenuation maps \citep[][]{Melchior00}, and is a
result of a very low dust column, with the SED fit returning an
average dust column density of only
$\sim4.5\times10^{4}M_{\odot}$\,kpc$^{-2}$ in the central region.

In contrast to the stellar mass, the bulge
contributes little to the total dust mass in M31. At larger radii,
emission from the disk of M31 along the minor axis overlaps with (and
dominates over) the dust emission from the bulge. Fits to the
integrated IR spectral energy distribution (SED) of the Andromeda
galaxy (within a 21.5 kpc aperture, Krause et al., in prep.), using
both the \citet{Draine07} and \citet{daCunha08} models, reveal that
the dust mass in the inner 2kpc only contributes $\sim$0.5\% to the
total dust mass of M31.  However, the central 1 kpc contributes a
ten-times larger fraction (i.e.~$\sim5$\%) to the total IR luminosity
of M31, because of the relatively warmer dust emission. We can
quantify the temperature of this dust in the bulge and in the rest of
M31 by fitting simple modified blackbodies.

\section{The dust heating in the M31 bulge}\label{sec:bulgeheat}

\subsection{Simple modified blackbodies}\label{sec:sBB}
We now convert the spatially resolved thermal-IR SEDs into dust
temperatures, describing the emission locally with
a single dust temperature, ${\rm T}_{\rm d}$. While a simplification,
this method returns a reasonable estimate for the mean, luminosity-weighted
temperature of the dust.

We do this by using 
a simple modified blackbody;
\begin{equation}\label{eqn:mBB}
S_{\nu}=N_{\rm d}\kappa_{\nu0}B_{\nu}({\rm T}_{\rm d})\left(\frac{\nu}{\nu_{0}}\right)^{\beta},
\end{equation}
where the dust surface brightness, $S_{\nu}$, is proportional
to the Planck function for the given dust temperature, $B_{\nu}({\rm
T}_{\rm d})$, modified by the dust emissivity, which is assumed to be a
power-law function of frequency,
$\kappa_{\nu0}\left(\nu/\nu_{0}\right)^{\beta}$. At a given dust
temperature this is then simply multiplied by the
dust column density, $N_{\rm d}$, to yield the surface brightness. 

The power-law function
is a reasonable approximation for the dust emissivity \citep[see
e.g.][]{Hildebrand83,Draine03}, with the galactic diffuse ISM well fit
by  an emissivity index of $\beta=2$ \citep{Draine84}. While a mean temperature for the dust can be
determined, in reality there will be a range of
temperatures along the lines-of-sight due to both a distribution of grain sizes and heating
radiation fields \citep[see, e.g.][]{Draine07}. This simplified
modelling foremost reflects a physically-motivated conversion of the FIR colours into the
three parameters returned by the fitting procedure: ${\rm T}_{\rm d}$,
$\beta$, and $N_{\rm d}\kappa_{\nu0}$. We refer to \citet{Shetty09a,Shetty09b}
for a more detailed description of the 
issues and uncertainties in using simple modified blackbodies for
integrated IR data.

We fit the modified blackbody to the 100-500\mum\ Herschel bands at
the 500\mum\ resolution. We do not include the PACS 70\mum\ data due
to the lower S/N across the M31 image, and as we
expect the emission at these wavelengths to have a significant
contribution from stochastically heated dust grains.
We limit our fitting to all pixels with $S/N>5$ in all 5 bands, which in practice is limited
predominantly by the PACS 100\mum\ band. To fit the simple modified blackbody to
the data, we assume a uniform, bounded
prior grid for all three parameters, compute the $\chi^2$ goodness of
fit for every model parameter set $j$,
\begin{equation}
\chi^2_{j}=\sum_{\nu}\left(\frac{S_{\nu,{\rm obs}}-S_{\nu,j}}{\sigma_{\nu}}\right)^2,
\end{equation}
where $S_{\nu,{\rm obs}}$ and $\sigma_{\nu}$ are the observed surface
brightness and observational uncertainty for the band $\nu$,
respectively, and $S_{\nu,j}$ is the model flux for parameter set,
$j$, determined from Equation \ref{eqn:mBB}. The bounds for
dust temperature and emissivity were $5 < T_{\rm d} < 50$ and $0.5 <
\beta < 3.5$, respectively. We then determine the
probability for each parameter set assuming a gaussian distribution, $\exp(-\chi_{j}^2/2)$, and
marginalise over the other parameters to determine the probability
distribution function (PDF) for each parameter, e.g.~T$_{\rm d}$. An
examination of the PDF for T$_{\rm d}$ in several pixels shows a
symmetric Gaussian distribution (as can be seen in the symmetric
uncertainties in Figure \ref{fig:m31Twrad}), providing justification
for our assumed Gaussian uncertainties for the parameters. 

\subsection{The dust temperature across M31}\label{sec:m31temp}
\begin{figure*}
\includegraphics[width=\hsize]{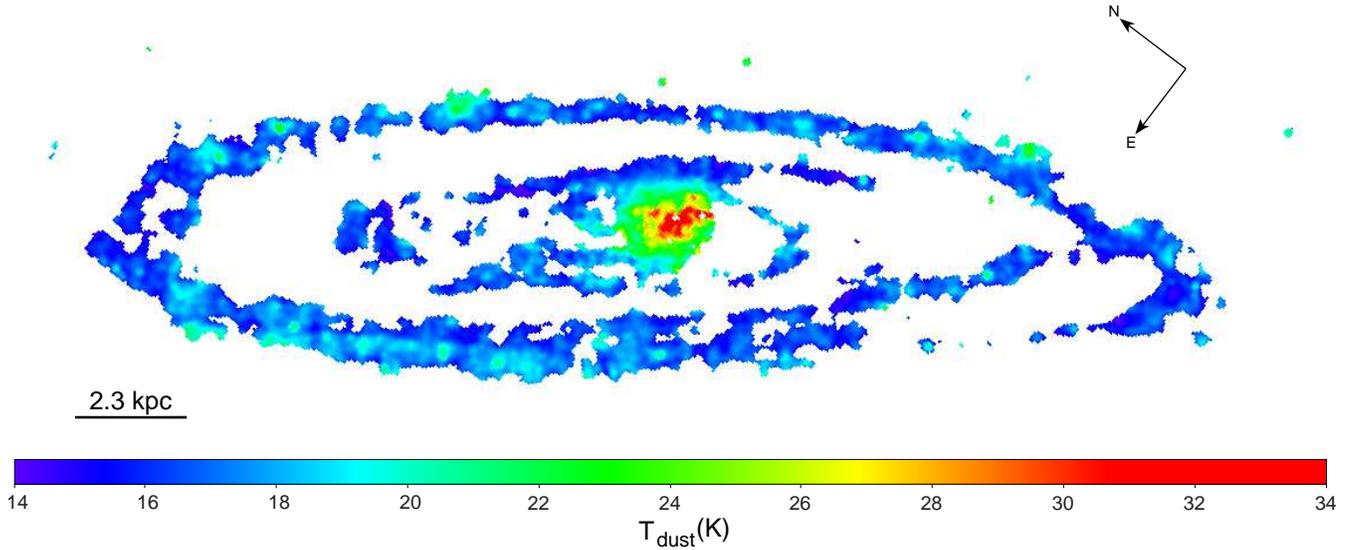}
\caption{Effective dust temperature map of M31, determined from a single
  modified blackbody fit to each matched pixel of the PACS100, 160\mum\ and
  SPIRE 250, 350, and 500\mum\ bands, with all bands convolved to
  the SPIRE 500\mum\ PSF and resolution (as described in sections
  \ref{sec:sBB} and \ref{sec:m31temp}). The median temperature is 17K, and is dominated
  by the emission  in the ring, while the highest temperature is
  reached in the nuclear region, as can be seen by the clear gradient
  in the figure. The median uncertainty is $\sim 1$K or $\sim$6\%.}\label{fig:m31temp} 
\end{figure*}

\begin{figure}
\includegraphics[width=\hsize]{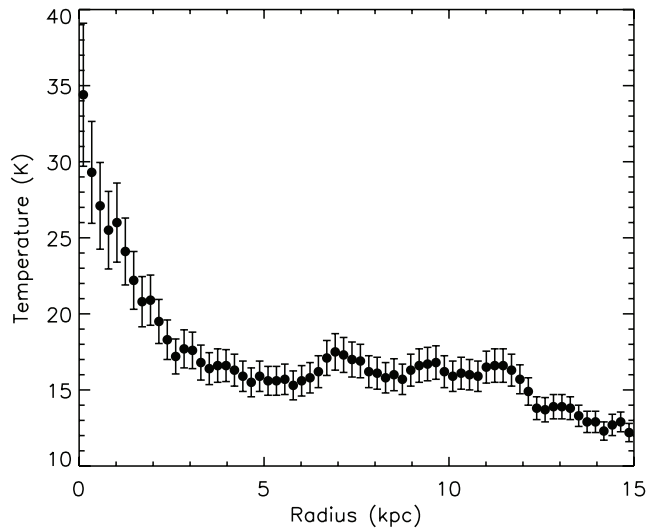}
\caption{Radial variation of mean temperature in M31. The temperature
  is determined by fitting the median flux in all bands in radial bins
  of 230pc width using elliptical annuli. The error bars
  show the 16-84 percentiles of marginalised PDFs.
}\label{fig:m31Twrad} 
\end{figure}

We performed this fit for every $14\asec$ pixel in the M31 IR data
that laid above our $S/N$ cut, with the resulting image for the dust
temperature shown in Figure \ref{fig:m31temp}. Pixels which fall below
our $S/N$ cut are shown as white, with the dust temperature (K) shown by
the colour, as indicated by the colour-bar at the bottom of the
image. Note that within 2 kpc, almost all pixels have sufficient S/N.
The median temperature across the image is 17K, close to MIPS based
disk temperature of $18\pm1$\,K determined by \citet{Tabatabaei10}.
The uncertainties (half
the 16-84 percentiles, or $1\sigma$ if the uncertainties are Gaussian) range from 5 to 15\%, with the largest
uncertainties in regions at the lowest signal to noise and the central
kiloparsec. The median uncertainty is $\sim$6\% ($\sim 1$K) dominated
by the emission in the dusty, star-forming ring seen in GALEX and FIR
images \citep[e.g.,][]{Thilker05}

 Two things stand out in this temperature map: the almost constant temperature
in the disk of M31, with T$_{\rm d}=17\pm1$K, and the strong temperature increase in
the central $\sim$2 kpc. The temperature in the 10kpc ring of M31 is
clearly not exactly constant, with warm spots occurring throughout the
ring. These warm spots correspond with the locations of H{\sc ii} regions in the
disk as determined by H$\alpha$ images \citep[c.f. the \ha\ maps of
][]{Devereux94,Azimlu11}.

\subsection{The radial dust temperature}\label{sec:radtemp}
The central dust temperature rise is more clearly quantified by taking the median
SED in radial bins. To do so we determined the median flux
in all bands in radial bins of 230 pc (using elliptical annuli with a
P.A.$=37.7^{\circ}$ and axis ratio of 0.26), and determined
the PDFs for all three parameters for each radial bin. The resulting
temperature variation with radius is shown in Figure
\ref{fig:m31Twrad}, with the 16--84 percentiles marked by the error
bars. This plot clearly shows the marked increase in dust temperature
in the inner 2kpc from the median disk dust temperature of T$_{\rm
  d}=16-17$\,K. The peak in dust temperature at the centre of $\sim35\,$K
is similar to the dust temperature of  T$_{\rm
  d}=33$\,K determined by \citet{Soifer86} for the central 4\amin, using IRAS data and
assuming a power law emissivity of $\beta=2$. Similarly,
\citet{Habing84} found 34K (using $\beta=1$) for approximately the same central
region and same data. As these works use apertures of
4\amin\ diameter, to define the
centre, these temperatures represent an average of the inner $\sim0.35-0.45$
kpc and thus are remarkably close to our determined value. As these
fits are based on only the IRAS 60 and 100\,\mum\ bands, and thus biased to
the warmer dust temperatures due to the shorter wavelengths, the
similarity in temperatures suggest 
that a single, warm component of dust dominates the IR SED in the
centre. This is supported by the similarity of the PACS 70\,\mum\ flux
and the model flux in the central kpc, even though not actually used
to determine the fit.

 In addition to giving access to the full IR SED, and
thus a better measure of dust temperature, the
higher resolution of \emph{Herschel} enables us 
to see the steep gradient in temperature from the disk to the
centre. This difference between the bulge and disk dust temperature
in M31 agrees with what \citet{Engelbracht10} found for the KINGFISH
sample of nearby galaxies \citep{Kennicutt11}.  \citet{Engelbracht10}
found that the ratio of  the central to disk dust temperature was
greater than 1 across their sample, and increased with increasingly
earlier types (see e.g.~their Figure 3).

\begin{figure}
\includegraphics[width=\hsize]{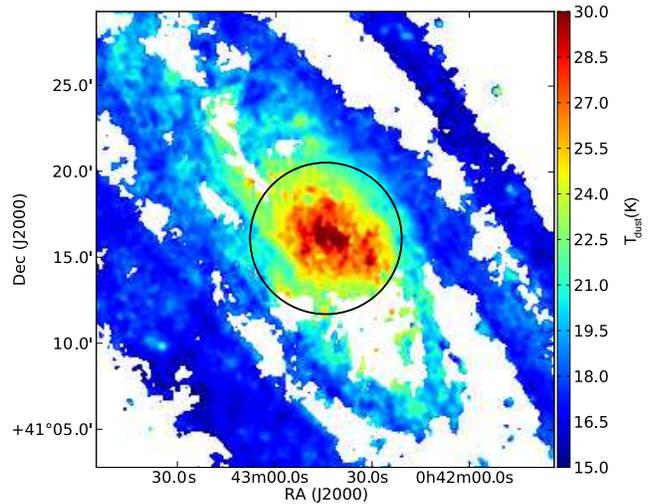}
\caption{Effective dust temperature map of the central region of M31, determined from a single
  modified blackbody fit to each matched 6\asec\ pixel of the convolved PACS100, 160 and
  SPIRE 250 \mum\ bands, assuming a constant emissivity of
  $\beta=2.0$. The black circle outlines the central kiloparsec region.
}\label{fig:m31T250}
\end{figure}

The emissivity, $\beta$, determined from these fits has a large
uncertainty across M31, and is consistent with $\beta=2$ at all
radii. Given this, we created a higher resolution temperature map
using only the 100, 160, and 250 \mum\ bands with the assumption  of a
constant emissivity of $\beta=2$. The resulting image of the central
region is shown in Figure \ref{fig:m31T250}, where the central
kiloparsec region is marked as a black circle and the median
uncertainty is 0.75 K. Interestingly, while more structure in
the temperature distribution in the centre is seen due to the higher resolution,
the overall temperature gradient in the centre still dominates the image. 
The steeper gradient in dust temperature along the
minor axis arises from the mixture of disk dust emission with
that from the bulge, leading to an overall lower average dust
temperature in these regions. The fine scale structure is likely due
to the distribution of the diffuse gas and dust as seen in Figure \ref{fig:m31ha_IR}.

\subsection{The heating mechanism of the bulge dust}

The clear temperature increase in the central regions of M31 as seen
in Figure \ref{fig:m31Twrad} is interesting as it closely corresponds
with the stellar light distribution as traced by the near-IR emission,
suggesting that the heating of the dust and bulge are linked.
As \citet{Li09} demonstrated (particularly
their Figure 3c), both the \ha\ and diffuse X-ray emission also
increase toward the centre in the same manner as the stellar light.
However, with the resolution offered by \emph{Herschel} we can begin
to understand the heating of the dust in the bulge of M31, and by
proxy in the bulges of early type spirals and in early-type galaxies.

\subsubsection{Potential heating sources}
In principle, one possibility for heating the dust is a low level of
star formation, buried amongst the dust and bulge stars. However, high resolution observations by HST
show that no OB stars are apparent throughout the bulge
\citep[e.g.][]{Brown98,Rosenfield12}, and no regions of high enough 
attenuation to bury such potential star formation are observed \citep{Melchior00}.  Also, the observed emission
line ratios in the ionized gas (e.g.~[\nii]/\ha) argue against young stars being the dominant
heating mechanism \citep{Ciardullo88,Saglia10}.

As the super massive black hole at the centre of M31 (dubbed M31$^*$) is
quiescent \citep{Li11}, heating by an AGN radiation field can be
discounted. In addition, the radial profile of dust temperature is incompatible with
a dominant central source heating, as shown in the following section.

Collisional heating of the dust by the hot $10^6$ K X-ray gas at the
centre of M31 \citep{Li07} is
another possibility \citep[see, e.g. ][]{Voit91,Natale10}. However,
the association of the dust with the \ha\ emission and the weak detection
of CO in regions of higher attenuation (discussed in section \ref{sec:dang})
indicate that the dust is in a denser medium, and not 
predominantly associated with the X-ray gas itself. Likewise, the total energy in
the diffuse X-ray gas is insufficient to fully explain the \ha\
emission, and thus is unlikely to heat the dust either
\citep{Li09}. Thus, while this may contribute to the emission, it is
insufficient to explain the majority of the hot dust emission.

\subsubsection{Bulge star heating}
This leaves the high stellar density of old stars at the
centre as the dominant heating mechanism of the dust in the bulge of
M31, as originally suggested by\citet{Habing84}.
By comparing the stellar light distribution for
the bulge against the distribution of temperatures in the central
kiloparsecs, we can now demonstrate the link between the heating of dust
by the bulge and the dust emission. In the case of a simple
steady-state temperature, grain cooling balances the dust heating, and
the grain temperature is then proportional 
to \citep[as shown in, e.g.][particularly Equations 24.17 and
24.18]{Draine11}; 
\begin{equation}\label{eqn:TdU1}
T_{\rm d}=\left(\frac{h\nu_{0}}{k_{\rm B}}\right)^{\beta/(4+\beta)}
\left[\frac{\pi^4\langle Q_{\rm abs}\rangle_{\star}c}{60\Gamma(4+\beta)\zeta(4+\beta)Q_{0}\sigma}\right]^{1/(4+\beta)}
 U_{\ast}^{1/(4+\beta)},
\end{equation}
where $Q_0$ and $\langle Q_{\rm abs}\rangle_{\star}$ are the dust
absorption cross-section at the reference frequency $\nu_0$ and
averaged across the heating spectrum \citep[Equation 24.2
in][]{Draine11}, respectively, $\Gamma$ and $\zeta$ are the Gamma and
Riemann zeta functions and  $U_{\ast}$ is the heating stellar
radiation field density. In the
simple assumption where the emissivity slope is constant at $\beta=2$, this reduces to
\begin{equation}\label{eqn:TdU2}
T_{\rm d} = \left(\frac{21 h^2c}{160 \pi^2k_{\rm B}^2\sigma}\right)^{\frac{1}{6} }
\left(\langle Q_{\rm abs}\rangle_{\star,0} \nu_0^2 U_{\ast}\right)^{\frac{1}{6} },
\end{equation}
where $\langle Q_{\rm abs}\rangle_{\star,0}$ is the
spectrally-averaged dust absorption cross-section normalised at $\nu_0$. Using a
reference wavelength of $\lambda_0=100$\,\mum\ and the unattenuated
spectral energy distribution determined in section \ref{sec:SED}, $\langle Q_{\rm
  abs}\rangle_{\star,0}\sim197$ for a Milky Way dust model with
$R_{\rm V}=3.1$\footnote{Available from
  \url{http://www.astro.princeton.edu/~draine/dust/dustmix.html}}
\citep[at $\lambda_0=250$\,\mum, $\langle Q_{\rm
  abs}\rangle_{\star,0}$ is approximately $(2.5)^2$ larger;][]{Draine03}.  

We can estimate the stellar radiation field heating the dust with a
few simple assumptions: {\bf 1)} the dust is in a thin disk at the
centre of M31 and {\bf 2)} is optically thin to the dominant heating radiation, and
{\bf 3)} the bulge stellar mass is spherically distributed and {\bf
  4)} has constant colours and stellar population. 
The first assumption is based on the findings given in section
\ref{sec:dang}, where the correlation of ionized gas and dust suggest
a thin disk geometry.
The measured opacity in the bulge of M31 is low, as shown by
\citet{Melchior00}, as is the average dust column of
$\sim4.5\times10^{4}M_{\odot}$\,kpc$^{-2}$ (see \S \ref{sec:SED}), thus the
assumption of optically thin dust is not unreasonable. 
As for the third assumption, the bulge of M31 has
already been shown
to be non-symmetric, displaying a boxy structure, and possibly even
different P.A.~and inclination angle to the disk
\citep[e.g.][]{Courteau11}. However, a simple spherical approximation for
the bulge with a constant mass-to-light ratio can actually reasonably
well reproduce the stellar light profile, as shown by
\citet{Geehan06}. The NIR colours are close to constant as shown in
Figure 1 in \citet{Courteau11}, but the UV colour, however, does show a gradient \citep[see,
e.g. Figure 3 in][]{Thilker05}. While some of this may arise from
a gradient in attenuation, there exists also radial gradients in the
UV-emitting stellar populations \citep{Rosenfield12} that may affect
our assumptions (though the follow section argues against this).

As shown in Appendix \ref{sec:ISRFb}, given these assumptions, using
the model for the bulge profile from \citet{Geehan06} and the
integrated unattenuated stellar luminosity from section \ref{sec:SED},
the bulge interstellar radiation field heating the dust is
\begin{equation}\label{eqn:Ub}
U_{\ast}(r)=\frac{3.66\times10^{-11}}{(r/{\rm kpc})}
\int_0^{\infty} \frac{{\rm ln}\left(\frac{r/r_{\rm b}+x}{|r/r_{\rm b}-x|}\right)}{(1+x)^3}dx 
\; {\rm erg\, cm^{-3},}
\end{equation}
at a radius $r$ (in kiloparsecs), and a bulge scale radius of $r_{\rm
  b}=0.61$\,kpc.

From Equations \ref{eqn:TdU2} and \ref{eqn:Ub}, we can then determine
the expected dust temperature (given the above assumptions) as a
function of spherical radius. For an infinitely thin disk, located in
the plane of M31, we can equate the dust temperature in spherical
coordinates with that in cylindrical (i.e.~$r=R$, $z=0$), giving
\begin{equation}\label{eqn:TdUR}
 T_{\rm d,U_{\ast}}(R)=1750\,\left[U_{\ast}(R)\right]^{\frac{1}{6}} \,{\rm K},
\end{equation}
where the factor 1750 arises from using the Milky Way dust model
described above for $\langle Q_{\rm abs}\rangle_{\star,0}$, and
$U_{\ast}$ is in erg\,cm$^{-3}$ as in Equation \ref{eqn:Ub}.  

\begin{figure}
\includegraphics[width=\hsize]{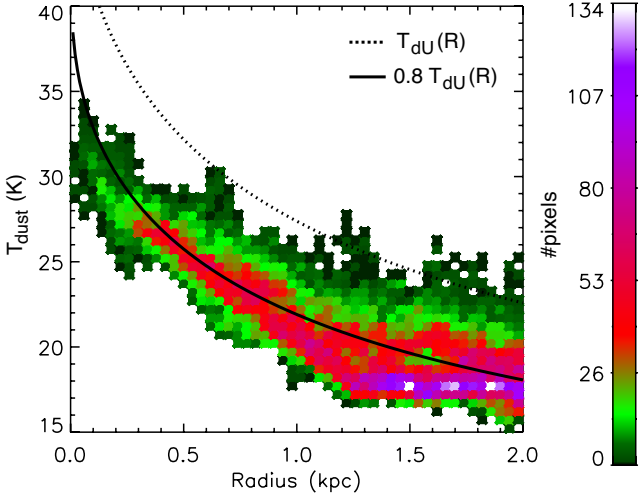}
\caption{The distribution of dust temperatures within the
  inner 2 kpc ($530\asec$) of M31. The colours show the number density
  of the $6\asec$ pixels from Figure \ref{fig:m31T250} in terms of
  $(T_{\rm d})$ and circular radius, as labelled by the colour bar to
  the right. Overlaid on this are two curves showing expected
  heating as based on the determined bulge ISRF. The dotted and solid
  curves show, respectively, the expected dust temperature distribution, 
  $T_{\rm d,U_\ast}(R)$, from Equation \ref{eqn:TdUR} and the same
  $T_{\rm d,U_\ast}(R)$ scaled by 0.8. 
}\label{fig:Tdist} 
\end{figure}

In Figure \ref{fig:Tdist} we show the distribution of the $6\asec$
pixels from the dust map in Figure \ref{fig:m31T250} in terms of the dust
temperatures with distance from the centre, based on our simple
circular radius. Overlaid on this distribution  are two curves. 
The expected dust heating distribution, $T_{\rm d,U_\ast}(R)$, as
given by Equation \ref{eqn:TdUR}, is shown by the dotted curve. 
As is clear from the figure, this is in excess of the observed
temperature in M31's bulge. The solid curve shows the same $T_{\rm
  d,U_\ast}(R)$, however multiplied by a factor 0.8. This brings it in
agreement with the observed temperature distribution except at radii
larger than $\sim1.2$\,kpc.  The first thing to take from these two
curves is that the bulge provides a more than sufficient interstellar
radiation field (ISRF$_{\rm b}$) to heat the dust, with expected
temperatures in excess of that observed. However, importantly, the
observed gradient in temperatures matches that expected from heating
the bulge. This leaves the question, why is the expected
temperature distribution in excess of that observed?

%
%
There are several possibilities given our simple assumptions. 
One possible explanation is
suggested by the flattening of the $T_{\rm dust}$ pixel distribution
in Figure \ref{fig:Tdist} at radii $>1.3$\,kpc. This flattening is most likely due to the
increasing disk contribution to the IR at these radii, which acts to
lower the mean line-of-sight dust temperature.
This is approximately
the radius where the stellar mass distribution, and therefore light
distribution, becomes disk dominated \citep[see,
e.g.][]{Geehan06,Courteau11}, and thus disk dust is likely also
dominating the determined dust temperature. As discussed in section
\ref{sec:m31temp} and shown in Figure \ref{fig:m31temp}, the mean dust
temperature across the disk is $\sim 17$\,K, which is also where the
pixel distribution flattens to in Figure \ref{fig:Tdist}. Figure
\ref{fig:m31T250} also gives some indication of the disk contribution
to the bulge temperature, as there is clearly a flatter gradient of
temperatures along the major axis as compared to the minor axis, which
should have a greater contribution of disk dust due to the high
inclination of M31. However, an examination of the observed IR
SED within the central kiloparsec indicates that it is well described
by a single black-body, as mentioned in section \ref{sec:radtemp}.
As a confirmation we also fitted the pixel-SEDs in the central region
with a combination of two modified-blackbodies. For both blackbodies,
we assumed an emissivity of $\beta=2$, and let the temperature of the
first component free, and that of
the second, cooler component constrained to 17\,K. A simple fit found that
the central kpc region is dominated by the warmer component, whose
temperature distribution is almost indistinguishable from Figure
\ref{fig:Tdist}. Thus, while it must occur at radii $>1$\,kpc, the
contribution of the disk cannot explain the offset between the
theoretical curve and the observational-based temperature.

Another possibility is that the dust in the bulge is not Milky Way like, and our
assumed $\langle Q_{\rm abs}\rangle_{\star,0}$ is incorrect. For
example, if we assume a Milky Way model of dust, but with $R_{\rm
  V}=5.5$ (using the \citep{Weingartner01} model opacity), $\langle
Q_{\rm abs}\rangle_{\star,0}$ reduces to $\sim155$ (normalised at
$\lambda_0=100$\,\mum, as above).  Such a possibility is reasonable,
given that the extreme environs of the bulge may act to alter the dust
size distribution, destroying small grains leading to a flatter
opacity.
However, given the index of $1/6$ in Eqn.~\ref{eqn:TdU2}, the offset
in temperature requires a significant change in dust opacity if this
alone causes this offset.
Similarly, our estimate of $U_{\ast}$ is also likely
incorrect given that the bulge is not spherical \citep{Courteau11},
and the dust is not in a perfectly thin disk. However, the index of
$1/6$ again requires our $U_{\ast}$ to be over estimated by
approximately an order of magnitude if it alone is incorrect.

More likely, it is a combination of these, plus the possibility of
some self-shielding by dust (which acts to reduce the $U_{\ast}$ seen
by the dust), which lead to the difference between the theoretical
temperature distribution and that determined from the modified
blackbody fit to the data. Yet it should still be noted, that these
are all needed to reduce the theoretical heating to that observed. The
bulge stellar radiation field provides more than sufficient energy to
heat the warm dust at the centre of M31.

%
As a final check that the bulge stars are dominating the heating, we
can apply the same methodology and reasoning as used for Eqn.~\ref{eqn:TdU2},
but assuming that the
radiation field heating the dust arises from a central point source (i.e.~AGN or nuclear star
cluster), meaning that $U \propto r^2$. However, when $T_{\rm
  d,U_\ast}(R)$ is determined for this radiation field
(similar to Eqn.~\ref{eqn:TdUR}), the slope does not show the same form as in
Figure \ref{fig:Tdist}, demonstrating that the heating radiation field must be
extended. Even allowing for a greater contribution of the disk
emission to approximate the observed slope, to match the point source $T_{\rm
  d,U_\ast}(R)$ to the measured $T_{\rm d}$ requires a central
luminosity that would be obvious in the optical (or X-ray) emission, which is
discounted by observations as already discussed \citep{Li09}.

\subsection{The heating of dust by old stars}
While Figure \ref{fig:Tdist} clearly links the bulge with the dust
emission, it does not reveal what exactly is heating the dust.
Typically, due to the steep wavelength dependence of the dust opacity,
the dust IR emission has been associated with UV light and hence star
formation. Typically, diffuse dust is considered to be heated by a
local interstellar radiation field (ISRF) or scaling thereof
\citep[see e.g. ][]{Draine07}. As this ISRF is typically based on the local ISRF
of \citet{Mathis83}, the UV light dominates the heating of dust
\citep[e.g. Figure 4 in][]{Mathis83}, and as the
local UV light is dominated by B stars, the diffuse dust emission
indirectly traces B stars and thus star formation on longer ($\sim
100$ Myr) timescales. As shown by \citet{Thilker05}, the UV light strongly peaks
at the centre of M31. However this UV light in the bulge has been
shown to be not associated with star-formation, but rather arises from
extreme horizontal branch stars \citep{Brown98, Rosenfield12}.

\begin{figure}
\includegraphics[width=\hsize]{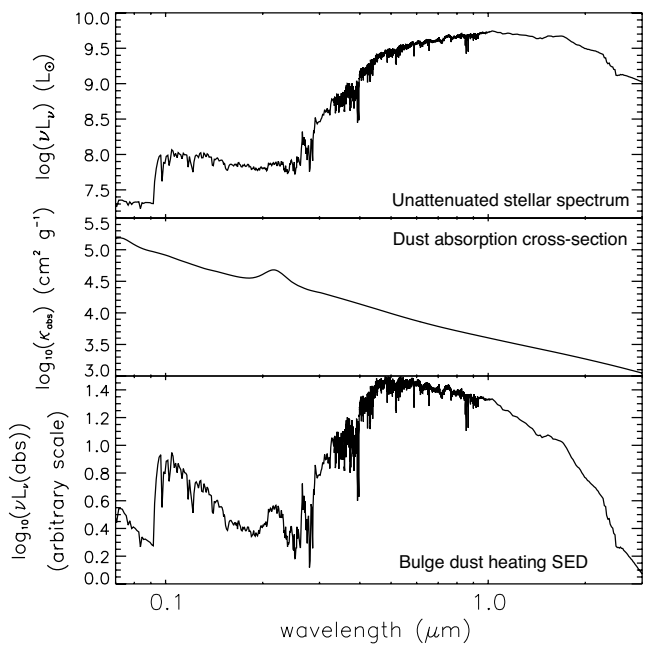}
\caption{The energy absorbed by diffuse dust in the bulge of M31. The
  top panel shows the ``unattenuated'' bulge UV--NIR SED from the
  MAGPHYS \citep{daCunha08} model fit to the integrated SED of the central
  kiloparsecs (blue curve in Figure \ref{fig:m31sed}). The middle
  panel shows the total absorption cross-section per gram of dust for
  the MW dust model with $R_{\rm V}=3.1$ from \citet{Draine03}. The
  bottom panel shows the product of these, revealing the wavelength
  distribution of the energy absorbed by dust. The conclusion is that
  the optical stellar light is the main contributor to the dust heating
  in the bulge. 
}\label{fig:bulge_dustabs}
\end{figure}

 However, while these hot, low-mass stars may be the source of the UV
upturn and possibly the \ha\ emission, they cannot be the dominant
heating source of the dust. Based on the IRAS and ANS observations,
\citet{Habing84} realised that the UV light would not be sufficient to
heat the dust, and the heating would be dominated by the $\lambda >300$ nm
light from evolved stars. 
In Figure \ref{fig:bulge_dustabs}, we show
three panels that illustrate which radiation heats the dust. In the
top panel we take the ``unattenuated SED'' from Figure
\ref{fig:m31sed}, which is likely an overestimate at
$\lambda<0.25\mu$m (see Section \ref{sec:FIR}). The middle panel shows
the total dust absorption cross-section per gram of dust for
the MW dust model with $R_{\rm V}=3.1$ from \citet{Draine03}, as
also used in the previous section. By multiplying these together we obtain the amount of
energy absorbed per H atom as a function of wavelength (bottom
panel). This Figure reveals that most of the energy that heats the
dust arises from optical photons, not the UV. As the $Q_{\rm
abs}(\nu)$ slope is $\approx 1$, the UV-optical slope needs to be
$<-1$ (in frequency) for optical photons to dominate the heating of
dust.

As mentioned before, the bottom panel of Figure
\ref{fig:bulge_dustabs} most likely overemphasises the role of UV photons in the
heating due to the small UV excess in the model stellar SED
(blue curve in Figure \ref{fig:m31sed}). The intrinsic UV emission produced by
the evolved stars in the centre of M31 (the so-called UV-upturn
observed in early-type galaxies) is well-reproduced by the
\citet{BC03} code, as shown by a comparison of models with the
observed UV colours of early-type galaxies \citep[see specifically
Figure 5 in][]{Donas07}. Even if the model UV is incorrect, the
intrinsic UV flux would still need to increase by $\sim 0.4$ dex (or
0.7 dex from the observed value) to be commensurate with the energy
deposited in the optical. This lack of UV heating is also clear when
the top panel of Figure \ref{fig:bulge_dustabs} is compared with the
ISRF of \citet{Mathis83} (e.g.~their Figure 1). When normalised to the peak at $\sim 1$\,\mum\ (in
both radiation fields), the UV in the IRSF is approximately an order
of magnitude greater than that determined in M31. 

Furthermore, the hot environment
of the circumnuclear region of M31 may destroy the smallest dust
grains as discussed in the previous section. The destruction of small grains flattens the
slope of $\kappa_{\rm abs}(\nu)$, leading to
yet a greater dominance of optical photons in the dust heating.  Thus
the bulge of M31 is a clear case where the heating of dust is
dominated by the light from old stars. This is in stark contrast to
typical star-forming galaxies, where the UV from young stars dominates
the heating of dust \citep[e.g.][]{Law11}. \citet{Monalto09} had
already found with GALEX--Spitzer data that the dust in the Andromeda
galaxy as a whole
appears to be heated mainly by stars a few Gyrs old. \citet{Habing84}
already inferred that  the
bulge of M31 takes this further with heating likely dominated by
evolved stars. However, the mean stellar age of  the bulge of M31 is
$>6$\,Gyr, with some estimates placing the dominant age $\gapprox
10$ Gyr \citep{Davidge05,Brown09,Saglia10}, demonstrating
that even the oldest stars are able to heat dust. This is a
clear cautionary note against using warm IR emission as a
direct tracer of star formation, or obscured UV emission.

\section{Discussion}\label{sec:disc}

As the nearest massive galaxy, the Andromeda galaxy allows us to
connect the small scale physics with the integrated properties of
galaxies. The central kiloparsec of M31 actually matches very well
many of the observed properties of 
nearby early-type galaxies. Within a 1
kiloparsec (4.4\amin) circular aperture the estimated stellar mass is
$\sim10^{10} M_{\odot}$, with a very old, red stellar light, estimated
to be greater than 10\,Gyrs old \citep{Saglia10}. The optical colours of
the bulge, e.g.~NUV-$r\approx5.0$, place it well in the realm of gas
poor early-type galaxies \citep[see, e.g. ][]{Oke68,Saintonge11,Smith11}.

There is a low level of dust attenuation across the centre, visible in
both the difference between the observed and modelled intrinsic SEDs
 in Figure \ref{fig:m31sed} and in the $A_{\rm B}$ map (their Figure 1)
 of \citet{Melchior00}. This
low level of attenuation is a result of the low dust column across the
centre and connected to the relatively weak IR emission.
 The total dust mass within this aperture
from the \citet{daCunha08} MAGPHYS model is $10^{5.2}M_{\odot}$,
contributing only 0.5\% of the total dust mass to M31. However, the
bulge contributes $\sim5$\% of the total IR luminosity due to the
relatively warm, blue FIR emission.  This dust mass results in a very
low $M_{\rm dust}/M_{\star}$ for the bulge, with $\sim
10^{-4.7}$. This actually places the bulge of M31 in a similar regime
as the sample of early-type galaxies explored in the \emph{Herschel}
reference survey \citep[HRS, ][]{Boselli10} in \citet{Smith11}, falling
somewhat in between the S0 and E galaxies (see their Figure 8).
Similarly, the temperature increase from disk to bulge in M31 follows the trend of having warmer dust in
earlier Hubble-types, as observed by \citet{Engelbracht10} in the 
KINGFISH sample. This trend was also seen  by \citet{Smith11} in the HRS,
who found warmer dust in E galaxies than
S0, and an overall warm dust temperature for the sample (mean $T_{\rm
  d} =24$K).  

The $M_{\rm gas}/M_{\star}$ ratio is also very low in the bulge of M31, with little
\hi\ and CO detected in the centre. However, ionized gas is seen in
\ha\, and the FIR emission is spatially well-correlated with this gas showing a
similar lower inclination, barred-spiral pattern as visible in Figure
\ref{fig:m31ha_IR}, with the MIR emission also following the diffuse
\ha\ morphology \citep{Li09}. Given this correlation, the \ha\ emission allows the
structure of the dust distribution to be determined
\citep[e.g.][]{Jacoby85}, and indicates that the ISM in the centre,
including the dust, is in a thin, spiral disk. The origin of this gas
and dust in the centre of M31 is still not known, and beyond the scope
of this work. \citet{Li09} suggest that the stellar ejecta
are more than sufficient to replenish the observed hot gas in M31's
centre, and suggest the rest of this matter comes out as a hot, X-ray
flow. However, the  spiral pattern observed in the bulge does appear
to link up with the emission in the disk (Figure \ref{fig:m31image}),
even if it appears to be at a different inclination to the disk.

Given the above similarities using the bulge of M31 as a resolved
representative of the spheroids of other early-type galaxies is a
reasonable assumption. Based on this and our resolved study of dust
heating in the bulge we can extrapolate to infer that a similar
heating mechanism for the dust 
occurs in the majority of early-type galaxies in which dust is
detected (i.e., the \citet{Smith11} sample).
While many early-type galaxies do have some AGN activity in
the centre, and several, generally with higher gas masses, have
observed active star formation \citep[e.g.][]{Crocker11}, the
combination of diffuse, optically thin dust and a strong, diffuse
radiation field due to the high stellar density is able to heat dust
at the centres of these early-type galaxies to a significantly warmer
level than that observed in most stellar disks. This increased dust
temperature makes the generally dust poor early-types still visible
in the recent and ongoing FIR surveys with \emph{Herschel}.

However, the possible contribution of disk dust to the observed IR
emission, as derived from the dust temperature offset in Figure
\ref{fig:Tdist}, suggests a final cautionary note in this
extrapolation. The bulge of M31 lies in a far different environment to
typical early-type galaxies, which affects both the observed emission, and
the evolution of the ISM in this early-type spheroid. 

\section{Summary}\label{sec:summ}
When observed in the far-infrared, the bulge of the Andromeda galaxy
(M31, NGC\,224) stands out as a region of luminous blue emission
(i.e. 70\mum\ bright), surrounded by ring-like red emission from the
disk.  This corresponds to a peak in the mean dust temperature,
T$_{\rm d}$ (Figure \ref{fig:m31temp}).  Across the disk of M31, the
mean dust temperature is reasonably constant with T$_{\rm
d}=17\pm1$K. However, within the central kiloparsec, the temperature
of the dust rapidly increases, reaching T$_{\rm d}\sim 35$K at the
centre, as seen in Figure \ref{fig:m31Twrad}.

The start of this upturn is also where the bulge begins to dominate
the stellar radiation field, clearly indicating that the heating
mechanism of this warm dust and the bulge are associated. Heating by
either an AGN or gas-grain collisions is discounted by X-ray
observations, leaving stellar heating as the likely source.  The
radial profile of this steep increase in temperature in the inner
kiloparsecs corresponds well with the temperature slope expected from
heating by the diffuse radiation field arising from the bulge stars,
suggesting direct link.  The theoretical dust temperature based on the
diffuse radiation field and our assumptions is actually in excess of
that observed, providing more than sufficient heating. This excess
suggests the not unreasonable possibilities of different dust
properties in the bulge from the standard diffuse dust in the Milky
Way and that the dust is likely to be in a clumpy distribution with
some self-shielding occurring.

However, even though
the bulge of M31 is observed in the UV, no young stars are seen in HST
observations or spectra \citep{Rosenfield12,Saglia10}, leaving old
($>6$ Gyr) stars as the dominant heating mechanism. Furthermore, by
taking the observed SED of the bulge and standard dust opacity it is
clear that it is the red optical light of these old stars that
dominates the heating of the warm dust in the centre, as originally
suggested by \citet{Habing84} with the IRAS observations.

Together this demonstrates the possibility of heating dust by stars
$>10$ billion years old. This is one of the clearest demonstrations of
how IR emission does not always correlate with star formation, a
common assumption, but rather is dependent upon the distribution of
dust (and associated gas) and the total radiation field, which
includes both young and old stars.  The bulge of M31 also demonstrates
one of the plausible mechanisms for heating the relatively warm dust
observed in the centres of early-type spirals, and early-type
galaxies. 

\section*{Acknowledgements}
The authors would like to thank P. Barmby and K. Gordon for providing
the \emph{Spitzer} IRAC and MIPS data for M31, respectively. B.G. would also
like to thank R. Shetty, A. Stutz, G. Seidel, and G. van der Ven for interesting and helpful
discussions. F.S.T. acknowledges the support by the DFG via the grant TA
801/1-1. This work made use
of the NASA Extragalactic Database, and the IDL astro library routines.

\bibliographystyle{mn2e}

\begin{thebibliography}{}
\bibitem[Aihara et al.(2011)]{Aihara11} Aihara, H., Allende 
Prieto, C., An, D., et al.\ 2011, \apjs, 193, 29 
%
\bibitem[Aniano et al.(2011)]{Aniano11} Aniano, G., Draine, 
B.~T., Gordon, K.~D., \& Sandstrom, K.\ 2011, \pasp, 123, 1218
%
%
%
\bibitem[Azimlu et al.(2011)]{Azimlu11} Azimlu, M., Marciniak, 
R., \& Barmby, P.\ 2011, \aj, 142, 139
%
%
\bibitem[Barmby et al.(2006)]{Barmby06} Barmby, P., et al.\ 2006, \apjl, 650, L45 
%
%
%
%
\bibitem[Boselli et al.(2010)]{Boselli10} Boselli, A., Eales, S., Cortese, L., et al.\ 2010, \pasp, 122, 261 
%
\bibitem[Braun et al.(2009)]{Braun09} Braun, R., Thilker, 
D.~A., Walterbos, R.~A.~M., \& Corbelli, E.\ 2009, \apj, 695, 937 
%
\bibitem[Brown et al.(1998)]{Brown98} Brown, T.~M., Ferguson, 
H.~C., Stanford, S.~A., \& Deharveng, J.-M.\ 1998, \apj, 504, 113
%
\bibitem[Brown(2009)]{Brown09} Brown, T.~M.\ 2009, Galaxy 
Evolution: Emerging Insights and Future Challenges, 419, 110 
%
\bibitem[Bruzual \& Charlot(2003)]{BC03} Bruzual, G., \& Charlot, S.\ 2003, \mnras, 344, 1000 
%
%
\bibitem[Cappellari et al.(2011)]{Cappellari11} Cappellari, M., 
Emsellem, E., Krajnovi{\'c}, D., et al.\ 2011, \mnras, 413, 813 
%
\bibitem[Ciardullo et al.(1988)]{Ciardullo88} Ciardullo, R., Rubin, 
V.~C., Ford, W.~K., Jr., Jacoby, G.~H., \& Ford, H.~C.\ 1988, \aj, 95, 438 
%
\bibitem[Coleman et al.(1980)]{Coleman80} Coleman, G.~D., Wu, 
C.-C., \& Weedman, D.~W.\ 1980, \apjs, 43, 393
%
\bibitem[Courteau et al.(2011)]{Courteau11} Courteau, S., Widrow, 
L.~M., McDonald, M., et al.\ 2011, \apj, 739, 20
%
\bibitem[Crocker et al.(2011)]{Crocker11} Crocker, A.~F., Bureau, 
M., Young, L.~M., \& Combes, F.\ 2011, \mnras, 410, 1197
%
\bibitem[da Cunha et al.(2008)]{daCunha08} da Cunha, E., Charlot, 
S., \& Elbaz, D.\ 2008, \mnras, 388, 1595 
%
\bibitem[Dalcanton et al.(2012)]{Dalcanton12} Dalcanton, J.~J., 
Williams, B.~F., Lang, D., et al.\ 2012, arXiv:1204.0010 
%
%
\bibitem[Davidge et al.(2005)]{Davidge05} Davidge, T.~J., Olsen, 
K.~A.~G., Blum, R., Stephens, A.~W., \& Rigaut, F.\ 2005, \aj, 129,
201 
%
\bibitem[Devereux et al.(1994)]{Devereux94} Devereux, N.~A., 
Price, R., Wells, L.~A., \& Duric, N.\ 1994, \aj, 108, 1667 
%
%
\bibitem[Draine \& Lee(1984)]{Draine84} Draine, B.~T., \& Lee, H.~M.\ 1984, \apj, 285, 89 
%
\bibitem[Draine(2003)]{Draine03} Draine, B.~T.\ 2003, \araa,41, 241 
%
\bibitem[Draine \& Li(2007)]{Draine07} Draine, B.~T., \& Li, A.\ 2007, \apj, 657, 810 
%
\bibitem[Draine(2011)]{Draine11} Draine, B.~T.\ 2011, Physics of 
the Interstellar and Intergalactic Medium by Bruce T.~Draine.~Princeton 
University Press, 2011.~ISBN: 978-0-691-12214-4,
%
\bibitem[Donas et al.(2007)]{Donas07} Donas, J., Deharveng, 
J.-M., Rich, R.~M., et al.\ 2007, \apjs, 173, 597
%
\bibitem[Evans et al.(2010)]{Evans10} Evans, I.~N., et al.\ 
2010, \apjs, 189, 37 
%
\bibitem[Engelbracht et al.(2010)]{Engelbracht10} Engelbracht, C.~W., Hunt, L.~K.,
Skibba, R.~A., et al.\ 2010, \aap, 518, L56 
%
\bibitem[Geehan et al.(2006)]{Geehan06} Geehan, J.~J., Fardal, 
M.~A., Babul, A., \& Guhathakurta, P.\ 2006, \mnras, 366, 996
%
\bibitem[Gil de Paz et al.(2007)]{GildePaz06} Gil de Paz, A., 
Boissier, S., Madore, B.~F., et al.\ 2007, \apjs, 173, 185
%
\bibitem[Gordon et al.(2006)]{Gordon06} Gordon, K.~D., et al.\ 2006, \apjl, 638, L87 
%
%
\bibitem[Haas et al.(1998)]{Haas98} Haas, M., Lemke, D., Stickel, M., Hippelein, H., Kunkel, M., Herbstmeier, U., \& Mattila, K.\ 1998, \aap, 338, L33 
%
\bibitem[Habing et al.(1984)]{Habing84} Habing, H.~J., Miley, 
G., Young, E., et al.\ 1984, \apjl, 278, L59
%
\bibitem[Hernquist(1990)]{Hernquist90} Hernquist, L.\ 1990, \apj, 
356, 359 
%
\bibitem[Hildebrand(1983)]{Hildebrand83} Hildebrand, R.~H.\ 1983, 
QJRAS, 24, 267
%
103 
%
Ferguson, A.~M.~N., et al.\ 2005, \apj, 634, 287 
%
%
\bibitem[Jacoby et al.(1985)]{Jacoby85} Jacoby, G.~H., Ford, H., 
\& Ciardullo, R.\ 1985, \apj, 290, 136 
%
%
\bibitem[Kennicutt(1998)]{Kennicutt98} Kennicutt, R.~C., Jr.\ 1998, \araa, 36, 189 
%
\bibitem[Kennicutt et al.(2011)]{Kennicutt11} Kennicutt, R.~C., 
Calzetti, D., Aniano, G., et al.\ 2011, \pasp, 123, 1347 
%
\bibitem[Law et al.(2011)]{Law11} Law, K.-H., Gordon, K.~D., 
\& Misselt, K.~A.\ 2011, \apj, 738, 124
%
\bibitem[Leroy et al.(2011)]{Leroy11} Leroy, A.~K., Bolatto, 
A., Gordon, K., et al.\ 2011, \apj, 737, 12
%
\bibitem[Li \& Wang(2007)]{Li07} Li, Z., \& Wang, Q.~D.\ 2007, \apjl, 668, L39 
%
\bibitem[Li et al.(2009)]{Li09} Li, Z., Wang, Q.~D., 
\& Wakker, B.~P.\ 2009, \mnras, 397, 148
%
\bibitem[Li et al.(2011)]{Li11} Li, Z., Garcia, M.~R., 
Forman, W.~R., et al.\ 2011, \apjl, 728, L10 
%
%
%
\bibitem[Mathis et al.(1983)]{Mathis83} Mathis, J.~S., Mezger, P.~G., \&
Panagia, N.\ 1983, \aap, 128, 212 
%
\bibitem[Melchior et al.(2000)]{Melchior00} Melchior, A.-L., 
Viallefond, F., Gu{\'e}lin, M., \& Neininger, N.\ 2000, \mnras, 312,
L29 
%
\bibitem[Melchior \& Combes(2011)]{Melchior11} Melchior, A.-L., \& Combes, F.\
2011, arXiv:1103.3392 
%
\bibitem[Montalto et al.(2009)]{Monalto09} Montalto, M., Seitz, S., Riffeser, A.,
et al.\ 2009, \aap, 507, 283 
%
\bibitem[Morganti et al.(2006)]{Morganti06} Morganti, R., de 
Zeeuw, P.~T., Oosterloo, T.~A., et al.\ 2006, \mnras, 371, 157 
%
\bibitem[Mutch et al.(2011)]{Mutch11} Mutch, S.~J., Croton, 
D.~J., \& Poole, G.~B.\ 2011, \apj, 736, 84
%
\bibitem[Natale et al.(2010)]{Natale10} Natale, G., Tuffs, 
R.~J., Xu, C.~K., et al.\ 2010, \apj, 725, 955 
%
\bibitem[Nieten et al.(2006)]{Nieten06} Nieten, C., Neininger, N., Gu{\'e}lin, M., et al.\ 2006, \aap, 453, 459 
%
\bibitem[Oke \& Sandage(1968)]{Oke68} Oke, J.~B., \& Sandage, A.\
1968, \apj, 154, 21 

\bibitem[Olsen et al.(2006)]{Olsen06} Olsen, K.~A.~G., Blum, 
R.~D., Stephens, A.~W., et al.\ 2006, \aj, 132, 271
%
%
%
\bibitem[Rich et al.(2005)]{Rich05} Rich, R.~M., Corsi, C.~E., 
Cacciari, C., Federici, L., Fusi Pecci, F., Djorgovski, S.~G., 
\& Freedman, W.~L.\ 2005, \aj, 129, 2670 
%
\bibitem[Rosenfield et al.(in prep.)]{Rosenfield12} Rosenfield, P. \& PHAT
  Team 2012, \apj, in prep.
%
\bibitem[Roussel(2012)]{Roussel12} Roussel, H.\ 2012, \aap, submitted
%
\bibitem[Rowlands et al.(2011)]{Rowlands11} Rowlands, K., Dunne, 
L., Maddox, S., et al.\ 2011, \mnras, 1996
%
%
\bibitem[Saglia et al.(2010)]{Saglia10} Saglia, R.~P., Fabricius, M., Bender, R., et al.\ 2010, \aap, 509, A61 
%
\bibitem[Saintonge et al.(2011)]{Saintonge11} Saintonge, A., 
Kauffmann, G., Kramer, C., et al.\ 2011, \mnras, 415, 32 
%
%
\bibitem[Shetty et al.(2009a)]{Shetty09a} Shetty, R., Kauffmann, 
J., Schnee, S., Goodman, A.~A., \& Ercolano, B.\ 2009a, \apj, 696, 2234 
%
\bibitem[Shetty et al.(2009b)]{Shetty09b} Shetty, R., Kauffmann, 
J., Schnee, S., \& Goodman, A.~A.\ 2009b, \apj, 696, 676 
%
s\bibitem[Smith et al.(2011)]{Smith11} Smith, M.~W.~L., Gomez, 
H.~L., Eales, S.~A., et al.\ 2011, arXiv:1112.1408 
%
\bibitem[Soifer et al.(1986)]{Soifer86} Soifer, B.~T., Rice, 
W.~L., Mould, J.~R., et al.\ 1986, \apj, 304, 651 
%
\bibitem[Stanek \& Garnavich(1998)]{Stanek98} Stanek, K.~Z., \& Garnavich, P.~M.\ 1998, \apjl, 503, L131 
%
\bibitem[Tabatabaei \& Berkhuijsen(2010)]{Tabatabaei10} Tabatabaei, F.~S., \& Berkhuijsen, E.~M.\ 2010, \aap, 517, A77 
%
\bibitem[Thilker et al.(2005)]{Thilker05} Thilker, D.~A., Hoopes, 
C.~G., Bianchi, L., et al.\ 2005, \apjl, 619, L67
%
%
\bibitem[Walterbos \& Kennicutt(1987)]{Walterbos87a} Walterbos, R.~A.~M., \& Kennicutt, R.~C., Jr.\ 1987, \aaps, 69, 311 
%
\bibitem[Weingartner \& Draine(2001)]{Weingartner01} Weingartner, J.~C., \& Draine, B.~T.\ 2001, \apj, 548, 296 
%
%
%
\bibitem[Voit(1991)]{Voit91} Voit, G.~M.\ 1991, \apj, 379, 122
%
%
%
\bibitem[Young et al.(2011)]{Young11} Young, L.~M., Bureau, M., 
Davis, T.~A., et al.\ 2011, \mnras, 414, 940
\end{thebibliography}

\appendix
\section{The bulge radiation field}\label{sec:ISRFb}
For the bulge radiation field we first need the stellar luminosity
distribution (which we label $\nu_{\star}(r)$). For this we follow
\citet{Geehan06} and use a \citet{Hernquist90} mass
profile due to its simple spherical geometry and assume a
constant mass-to-light ratio, giving;
\begin{equation}\label{eqn:nustar}
\nu_{\star}(r)=
\frac{L_{\rm b}}{2\pi r_{\rm b}^3}\frac{1}{(r/r_{\rm b})(1+r/r_{\rm b})^3}.
\end{equation} 
\citet{Geehan06} found the bulge radius to be 0.61 kpc (their Equation 1 and Table 2), and we used
the determined unattenuated stellar luminosity within 1 kpc of
$L_{\star} = \sim10^{9.9}\,L_{\odot}$ (from section \ref{sec:SED}) to
normalise $L_{\rm b}$ to $10^{10.32}L_{\odot}$.

The interstellar radiation field in the bulge (ISRF$_{\rm b}$) is then simply the
luminosity density (in e.g.~$L_{\odot}$\,pc$^{3}$) convolved with
spherical dilution, $1/r^2$;
\begin{eqnarray}\label{eqn:conv}
U_{\ast}(r)&=&\nu_{\star}(r)\ast\frac{1}{4\pi r^{2}c}\nonumber \\
 &=&\int_V \nu_{\star}(r_0)\frac{1}{4\pi (\vec{r}_0 - \vec{r})^{2}c} dV,
\end{eqnarray}
where the factor $1/c$ is introduced to convert the ISRF$_{\rm b}$ to
an energy density (i.e.~erg cm$^{-3}$). As both the luminosity density
and dilution are purely radial functions, the volume integral in Equation \ref{eqn:conv}
reduces to (in spherical coordinates); 
\begin{eqnarray}\label{eqn:conv2}
U_{\ast}(r)&=& \int_0^{\infty} \frac{\nu_{\star}(r_0)}{4\pi c}
\int_0^{2\pi}\int_0^{\pi} \frac{ r_0^2\sin \phi}{r_0^2-2rr_0\cos \phi +
  r^2}d\phi d\theta  dr_{0} \nonumber \\
&=& \int_0^{\infty} \nu_{\star}(r_0)\,\frac{r_0}{2rc}{\rm ln}\left(\frac{r+r_0}{|r-r_0|}\right) \,dr_{0}.
\end{eqnarray}

Substitution of Eqn.~\ref{eqn:nustar} into this and condensing then gives 
\begin{equation}\label{eqn:Ustar}
U_{\ast}(r)=\frac{L_{\rm b}}{4\pi r_{\rm b}c}\,\frac{1}{r}\,
\int_0^{\infty} \frac{{\rm ln}\left(\frac{r/r_{\rm b}+x}{|r/r_{\rm b}-x|}\right)}
 {(1+x)^3}dx, 
\end{equation}
where $x$ has been substituted for $r_0/r_{\rm b}$. The factor in
front of $1/r$ and the integral equals $3.66\times
10^{-11}$\,erg\,cm$^{-3}$ for our given $L_{\rm b}$ and $r_{\rm b}$,
with $r$ in kiloparsecs. While the integral in Eqn.~\ref{eqn:Ustar}
can be analytically determined, we use a numerical integration for
simplicity within this work. 

\end{document}